\documentclass[conference]{IEEEtran}
\IEEEoverridecommandlockouts
\usepackage{cite}
\usepackage{amsmath,amssymb,amsfonts}
\usepackage{algorithmic}
\usepackage{graphicx}
\usepackage{textcomp}
\usepackage{xcolor}
\usepackage{cite}

\usepackage{dsfont}
\usepackage{array}
\usepackage{url}
\usepackage{booktabs}
\usepackage{multirow}
\usepackage{caption}

\begin{document}
\title{Energy-Efficient Subchannel and Power Allocation for HetNets Based on Convolutional Neural Network}

\author{Di Xu$^{\dagger}$, Xiaojing Chen$^{\dagger}$, Changhao Wu$^{\dagger}$, Shunqing Zhang$^{\dagger}$,  Shugong Xu$^{\dagger}$ and Shan Cao$^{\dagger}$\\
$^{\dagger}$ Shanghai Institute for Advanced Communication and Data Science, \\
Key laboratory of Specialty Fiber Optics and Optical Access Networks, \\
Joint International Research Laboratory of Specialty Fiber Optics and Advanced Communication, \\
Shanghai University, Shanghai, 200444, China\\
Email:\{danny\_xd, jodiechen, wuchanghao, shunqing, shugong, cshan\}@shu.edu.cn}

\maketitle
\begin{abstract}
Heterogeneous network (HetNet) has been
proposed as a promising solution for handling the wireless traffic explosion in future fifth-generation (5G) system. In this paper, a joint subchannel and power allocation problem is formulated for HetNets to maximize the energy efficiency (EE). By decomposing the original problem into a classification subproblem and a regression subproblem, a convolutional neural network (CNN) based approach is developed to obtain the decisions on subchannel and power allocation with a much lower complexity than conventional iterative methods. Numerical results
further demonstrate that the proposed CNN can achieve similar performance as the Exhaustive method, while needs only 6.76\% of its CPU runtime.
\end{abstract}


\section{Introduction}
With the development of smart phones and wearable devices in the past decades, the volume of mobile traffic in communication networks has grown exponentially. Developing effective resource allocation schemes becomes increasingly crucial. Extensive studies have been carried out to
develop resource allocation schemes in various wireless networks.

In \cite{Shi2011An}, the authors designed transmit beamformers to maximize the sum-utility of a MIMO broadcast channel. Energy-efficient resource allocation was investigated for the uplink of multi-user multi-channel Orthogonal Frequency Division Multiplexing (OFDM) based systems \cite{Khakurel2012Energy}, and the heterogeneous networks (HetNets) \cite{Richter2009Energy,Tang2017Energy,Richter2010Micro,Zhang2017Downlink}.

Most previous works \cite{Khakurel2012Energy,Richter2009Energy,Tang2017Energy,Richter2010Micro,Zhang2017Downlink} derived the resource allocation strategies as the solutions of optimization problems, where iterative algorithms are applied, such as weighted minimum mean square error (WMMSE) in \cite{Shi2011An}. In iterative schemes, a large number of iterations need to be carried out before convergence is achieved. The high computational cost prevents implementing these algorithms in real-time for practical uses.

As a key technology in artificial technology, deep learning has been recently used for solving traditional problems in wireless communications, such as Polar decoding\cite{Gruber2017On, Cammerer2017Scaling} and Massive MIMO channel estimation\cite{He2018Deep, Wen2017Deep}. Deep neural networks (DNNs) can be used to solve complex nonlinear non-convex problems without constructing complicated mathematical models~\cite{Sun2017Learning,Lee2018Deep,Wang2018A}. For example, the work in \cite{Sun2017Learning} showed that DNN could be used to approximate the WMMSE proposed in \cite{Shi2011An}, with a much lower computational time.

In this paper, we propose a convolutinoal neural network (CNN) based resource allocation approach for HetNets~\cite{Agiwal2017Next, Zhang2015Cloud}.
The main contributions of this work can be summarized as follows.
\begin{itemize}
    \item Considering an OFDM-based HetNet, we formulate the resource allocation task as a joint subchannel and power allocation problem, which maximizes the energy efficiency (EE) of the network while satisfying the requirement of the spectrum efficiency (SE).
    \item Different from \cite{Sun2017Learning,Lee2018Deep,Wang2018A}, which either solve a regression problem or a classification problem for resource allocation by deep learning, the proposed CNN, for the first time, decomposes the original problem into a classification subproblem and a regression subproblem, to infer the energy-efficient decisions on joint subchannel and power allocation.
    \item Extensive numerical experiments are conducted to demonstrate that the proposed CNN can achieve similar performance as the Exhaustive method, while substantially reduce the computational time.
\end{itemize}

The rest of the paper is organized as follows. In Section~\ref{sect:pro}, we describe the system model and give the problem formulation. The proposed neural network architecture is developed in Section~\ref{sect:dnn}. Simulation results are provided in Section~\ref{sect:sim}, followed by the conclusion in Section~\ref{sect:conc}.

\section{System Model and Problem Formulation} \label{sect:pro}
\subsection{System Model}
    \begin{figure}[ht]
    \centering
    \includegraphics[width=8.5cm]{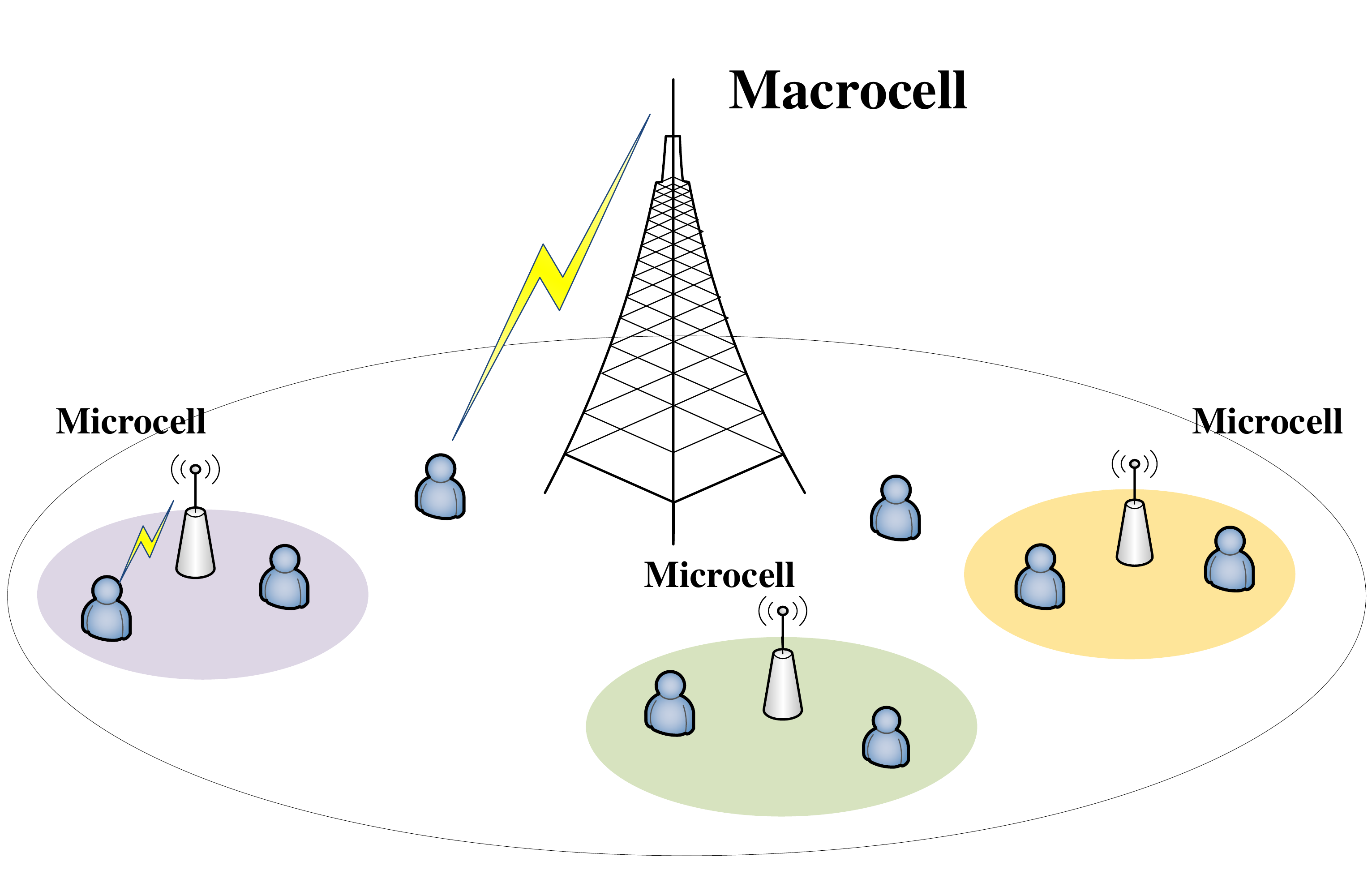}
    \caption{The Structure of an OFDM-based HetNet.}
    \label{User distrbution}
\end{figure}

Consider the downlink transmission in an OFDM-based HetNet, where a set $\mathcal{N}:=\{1,2,\ldots,N\}$ of BSs (i.e., macrocell and microcell BSs) serve a set $\mathcal{U}:=\{\mathcal{U}_{1},\mathcal{U}_{2},\ldots,\mathcal{U}_{N}\}$ of users; see Fig.~\ref{User distrbution}. Let $\mathcal{U}_{n}=\{1,2,\ldots,U_{n}\}$ denote the set of users communicating with BS $n$. Let $\mathcal{M}:=\{1,2,\ldots,M\}$ represent the set of macrocell BSs. The set of microcell BSs can be then given by $\mathcal{S}:=\{\mathcal{N}-\mathcal{M}\}$.

Each BS has a set $\mathcal{K}=\{1,2,\ldots,K\}$ of subchannels.
We define $l^{n}_{u,k}\in\{0,1\}$ as the subchannel allocation indicator. Let $l^{n}_{u,k}=1$ if subchannel $k$ is allocated to user $u$ by BS $n$; and $l^{n}_{u,k}=0$, otherwise.
Assume that each subchannel of a BS can be assigned to at most one user in its cell, and each user must get at least one subchannel, as given by

\begin{subequations}\label{subchannel_allocation}
\begin{align}
\sum\limits_{u\in\mathcal{U}_{n}}l^{n}_{u,k}\leq 1, \quad\forall{n}\in\mathcal{N}, {k}\in\mathcal{K}.\\
\sum\limits_{k\in\mathcal{K}}l^{n}_{u,k}\geq 1, \quad\forall{n}\in\mathcal{N}, {u}\in\mathcal{U}.
\end{align}
\end{subequations}

Let $p^{n}_{k}$ denote the transmit power of BS $n$ on subchannel $k$. We have
\begin{subequations}\label{powe_constraints}
\begin{align}
&0\leq \sum\limits_{k\in\mathcal{K}}p^{m}_{k}\leq P^{\mathcal{M}}_{\max},  \quad\forall{m}\in\mathcal{M}, \\
&0\leq \sum\limits_{k\in\mathcal{K}}p^{s}_{k}\leq P^{\mathcal{S}}_{\max},  \quad~\forall{s}\in\mathcal{S}, \\
&\quad~~ P^{\mathcal{M}}_{\max}>P^{\mathcal{S}}_{\max},\quad\quad\forall {m}\in\mathcal{M}, {s}\in\mathcal{S}, \label{larger_power}
\end{align}
\end{subequations}
where $P^{\mathcal{M}}_{\max}$ is the maximum transmit power of a macrocell BS and $P^{\mathcal{M}}_{\max}$ is the maximum transmit power of a microcell BS.
\eqref{larger_power} indicates that the maximum transmit power of a macrocell BS is larger than that of a  microcell one.

Let $h^{n}_{u,k}$ denote the channel gain from BS $n$ to user $u$ on subchannel $k$.
We assume that the BSs and users have perfect channel state information (CSI). Then the achieved transmit rate of user $u\in\mathcal{U}_{n}$ on subchannel $k$ is
\begin{equation}
   r^{n}_{u,k}=B\log_{2}(\frac{h^{n}_{u,k}p^{n}_{k}}{I^{n}_{u,k}+\sigma^{2}})
\end{equation}
where $B$ is the subchannel bandwidth, $I^{n}_{u,k}=\sum\limits_{j\in\mathcal{N} j\neq n}h^{j}_{u,k}p^{j}_{k}$ is the inter-cell interference, and $\sigma^{2}$ is the power of the Additive White Gaussian Noise (AWGN).
The overall throughput of the network is therefore given by
\begin{equation}
R(\boldsymbol{l},\boldsymbol{p})=\sum\limits_{n\in\mathcal{N}}\sum\limits_{u\in\mathcal{U}_{n}}\sum\limits_{k\in\mathcal{K}}l^{n}_{u,k}r^{n}_{u,k},
\end{equation}
where $\boldsymbol{l}:=\{l^{n}_{u,k}, \forall n,u,k\}$ and $\boldsymbol{p}:=\{p^{n}_{k}, \forall n,k\}$.
The total power consumption of the network can be described as
\begin{equation}
   p_{\text{tot}}(\boldsymbol{p})=\frac{1}{\rho}\sum\limits_{n\in\mathcal{N}}\sum\limits_{k\in\mathcal{K}}p_k^{n}+(Mp^{m}_{c}+Sp^{s}_{c}),
\end{equation}
 where $\rho$ is the power-amplifier inefficiency factor of BSs, and $p^{m}_{c}$ and $p^{s}_{c}$ are constant values denoting the circuit power consumption of macrocell BSs and microcell BSs, respectively. The system EE is defined as the ratio of achievable throughput to total power consumption in the HetNet, and the system SE is defined as the ratio of achievable throughput to system bandwidth, as given by
\begin{equation}
   \eta_{\text{EE}}(\boldsymbol{l},\boldsymbol{p})=\frac{R(\boldsymbol{l},\boldsymbol{p})}{ p_{\text{tot}}(\boldsymbol{p})},
\end{equation}
\begin{equation}
   \eta_{\text{SE}}(\boldsymbol{l},\boldsymbol{p})=\frac{R(\boldsymbol{l},\boldsymbol{p})}{KB}.
\end{equation}
With $\varepsilon$ denoting the target value of the system SE, it is required that
\begin{equation}\label{SE}
  \eta_{\text{SE}}(\boldsymbol{l},\boldsymbol{p})\geq\varepsilon.
\end{equation}

\subsection{Problem Formulation}
Our objective is to maximize the EE of the HetNet by making optimal decisions on the subchannel assignment $\boldsymbol{l}$ and power allocation $\boldsymbol{p}$, while guaranteeing the target SE. The problem of interest is to solve
\begin{align}\label{P1}
&\max\limits_{\{\boldsymbol{l},\boldsymbol{p}\}} \quad\quad\eta_{\text{EE}}(\boldsymbol{l},\boldsymbol{p})\\
&\text{s.t.} \quad \eqref{subchannel_allocation},~\eqref{powe_constraints}, \text{and}~\eqref{SE}.\notag
\end{align}

Note that problem \eqref{P1} is a mixed-integer nonlinear non-convex optimization problem, which is in general difficult to solve. In this paper, we propose a CNN-based approach to solve this problem.

\section{Machine learning for resource allocation} \label{sect:dnn}
In this section, we introduce our data generation process and give details on the CNN structure we propose.

\subsection{Data Generation}
In order to generate our data set, we set a HetNet in urban area scenario, which consists of one macrocell and several microcells. The specific setting of the HetNet will be given in Section IV. The channel gains $\{h_{u,k}^n,\forall u,k,n\}$ are first generated following a standard normal distribution, i.e., Rayleigh fading distribution with zero mean and unit variance. With fixed $P^{\mathcal{M}}_{\max}$, $P^{\mathcal{S}}_{\max}$ and $\sigma$, we generate the corresponding subchannel allocation indicator $\boldsymbol{l}$ and the allocated power $\boldsymbol{p}$ for each channel realization by running an exhaustive method. The Exhaustive method iteratively calculates and compares the EE for all possible schemes and chooses one of the scheme that maximizes the EE as the optimal solution. By doing so, the Exhaustive method sets a benchmark for the proposed CNN-based approach with a high complexity. By repeating the above process for a large number of times, we generate the entire training data set $\{h_{u,k}^n, \boldsymbol{l},\boldsymbol{p}\}$. Let a matrix $\mathbb{H}_{N\times U\times K}$ collect the channel gains from BS $n$ to user $u$ on subchannel $k$, i.e., $\{h_{u,k}^n,\forall u,k,n\}$.

\subsection{Proposed Convolutional Neural Network}
\begin{figure}[ht]
    \centering
    \includegraphics[width=8.5cm]{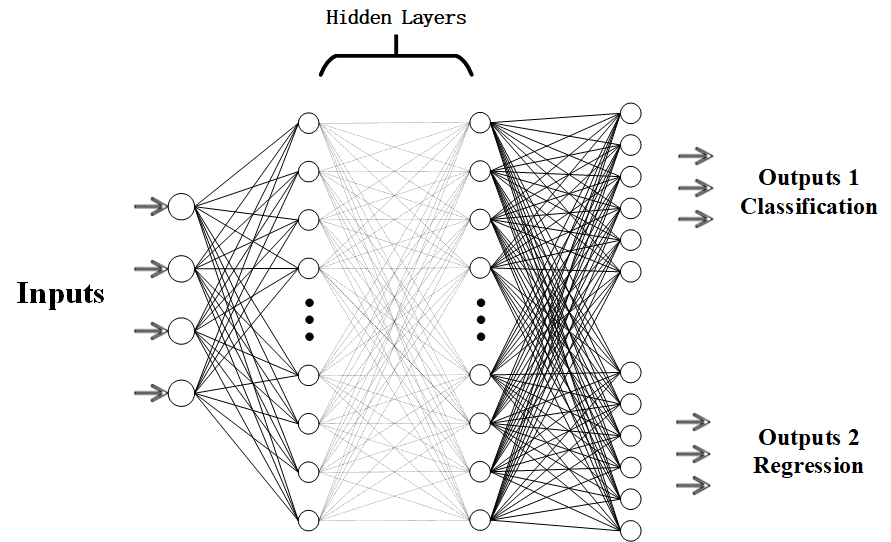}
    \caption{The CNN structure used in this work, which consists of one input layer, multiple hidden layers, and one output layer. The hidden layers are composed of four convolutional layers and three fully connected (FC) layer.}
    \label{fig:ML}
\end{figure}
Different from existing works which either solve a regression problem \cite{Sun2017Learning,Lee2018Deep}, or a classification problem~\cite{Wang2018A} for resource allocation by deep learning, our proposed CNN architecture decomposes the original problem into a classification subproblem and a regression subproblem, and then outputs the energy-efficient decisions on joint subchannel and power allocation. It consists of three part: input layer, hidden layers and output layer.

\begin{itemize}
  \item{\em \textbf{Input Layer}}: The input data is a three-dimensional matrix $\mathbb{H}_{N\times U\times K}$ collecting the channel gains from BS $n$ to user $u$ on subchannel $k$.
  \item{\em \textbf{Hidden Layers}}: The hidden layers are composed of four convolutional layers and three fully connected (FC) layers with the activation function, Rectified Linear Unit (ReLU). The reason a CNN is chosen as our neural network is that the sliding window of CNN can extract the features between the elements of the input matrix, which leads to better performance of classification and regression than other neural networks (e.g., a FC DNN). The parameters of the hidden layers will be given in Section~\ref{sect:sim}.
  \item{\em \textbf{Output Layer}}: Two sets $\widehat{\boldsymbol{l}}$ and $\widehat{\boldsymbol{p}}$ are output from this layer. Here, $\widehat{\boldsymbol{l}}$ is the subchannel allocation indicator that determines the allocation of subchannels to users; $\widehat{\boldsymbol{p}}$ collects the power allocation decisions that maximize the system EE. For the output  $\widehat{\boldsymbol{l}}$ applying to a classification subproblem, we choose \emph{linear} as the activation function; while for the output $\widehat{\boldsymbol{p}}$ applying to a regression subproblem, \emph{softmax} is selected as the activation function.
\end{itemize}

We use the training data set to optimize the weights of the CNN. The CNN is trained to regenerate the subchannel and power allocation derived from the Exhaustive method, given channel gains $\mathbb{H}_{N\times U\times K}$.
Since the proposed CNN aims to solve different subproblems (i.e., classification and regression), different loss functions are chosen adapting to different features of the subproblems.
\begin{itemize}
    \item {\em \textbf{Categorical Crossentropy}}
        \begin{equation}
            \mathbf{L_{reg}}( \boldsymbol{l},\widehat{ \boldsymbol{l}})=\sum\limits_{i}l_{i}\log(\widehat{l_{i}})
        \end{equation}
    where $\widehat{\boldsymbol{l}}$ is the predicted subchannel allocation indicator output by the CNN, and $\boldsymbol{l}$ is the subchannel allocation indicator in the training set. In (10), $\widehat{l_{i}}$ and $l_{i}$ are the elements in $\widehat{\boldsymbol{l}}$ and $\boldsymbol{l}$, respectively.

    \item {\em \textbf{Mean Square Error}}
        \begin{equation}
            \mathbf{L_{cls}}( \boldsymbol{p},\widehat{ \boldsymbol{p}}) = \sum\limits_{i}(p_{i}-\widehat{p}_{i})^{2}
        \end{equation}
    where $\widehat{\boldsymbol{p}}$ is the predicted power vector output by the CNN, and $\boldsymbol{p}$ is the allocated power in the training set. Also, $\widehat{p_{i}}$ and $p_{i}$ are the elements in $\widehat{\boldsymbol{p}}$ and $\boldsymbol{p}$, respectively. 
\end{itemize}
Therefore, the CNN is trained to minimize the following total loss function:
\begin{equation}
    \mathbf{L_{total}}=\mathbf{L_{reg}}( \boldsymbol{l},\widehat{\boldsymbol{l}})+\mathbf{L_{cls}}( \boldsymbol{p},\widehat{ \boldsymbol{p}})
\end{equation}

At last, CNN would be convergence by training process. The network parameters are shown in Section~\ref{sect:sim}.

\section{Simulation Results} \label{sect:sim}
\subsection{Simulation Configuration}
\begin{figure}[t]
\centering
\includegraphics[height=8cm,width=8.5cm]{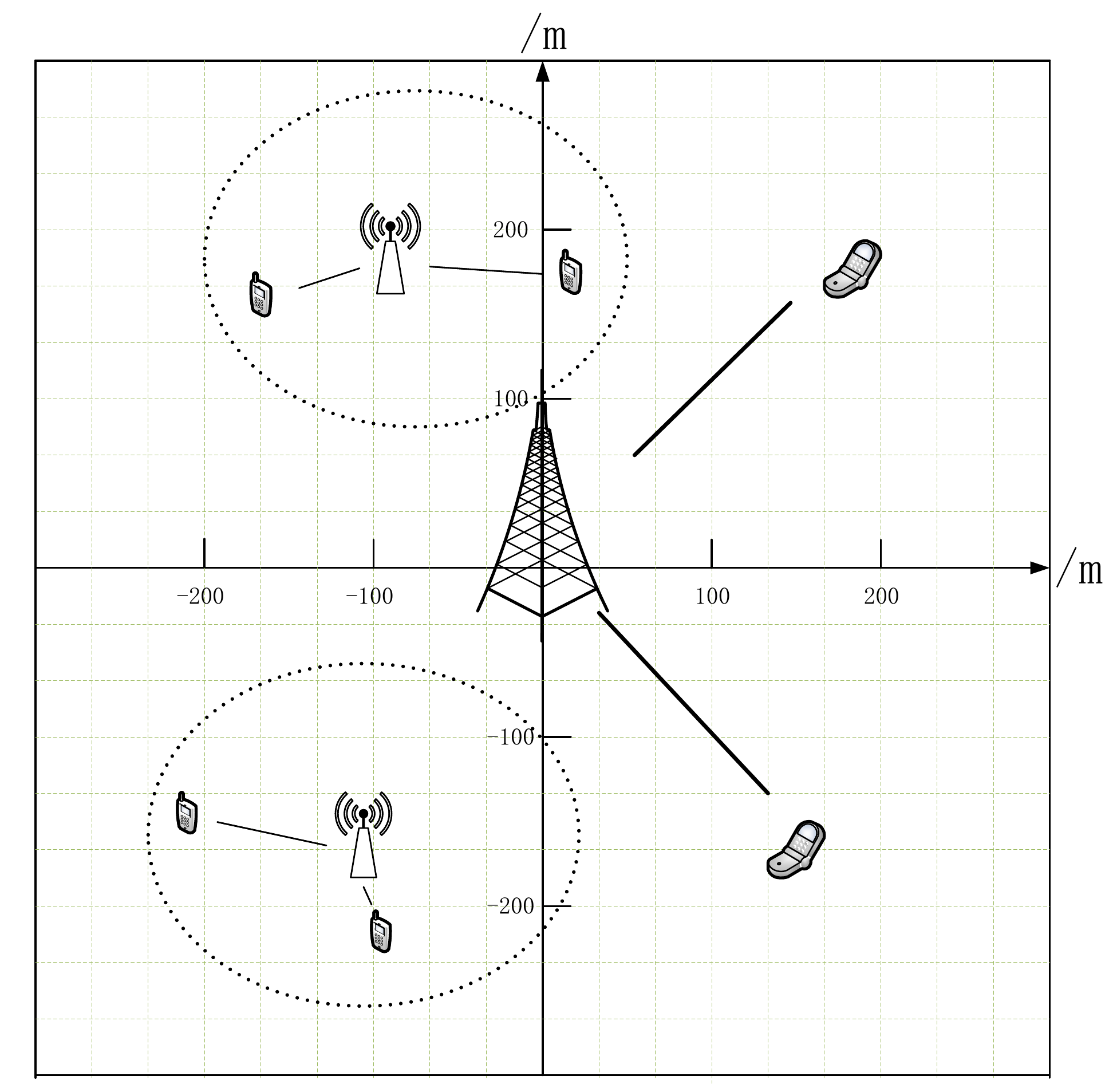}
\caption{Simulation model.}
\label{system model}
\end{figure}

In this section, we evaluate the performance of the proposed CNN approach. We consider a scenario where there is one macrocell and two microcells, each with a BS communicating with $U_n=2$ users; $K=2$ subchannels are allocated to each BS. The users are located uniformly in the entire cells; see Fig.~\ref{system model}. Table~\ref{net_par} summarizes the parameters of the HetNet scenario.

\begin{table}[ht]
    \caption{Parameters\label{net_par}}
    \centering
    \footnotesize
    \renewcommand\arraystretch{0}
    \setlength{\tabcolsep}{5.0mm}{
    \begin{tabular}{c c }
        \toprule
        Parameter             &Value\\
        \midrule
        \midrule
        System bandwidth       & 2 MHz\\
        \midrule
        Carrier frequency      & 2 GHz\\
        \midrule
        Number of total users     & 6 \\
        \midrule
       Number of macrocell BSs     & 1 \\
        \midrule
        Number of microcell BSs     & 2 \\
        \midrule
        Number of subchannels     & 2 \\
        \midrule
        Antenna height         & 15 m\\
        \midrule
        Macrocell pathloss     &  $128.1+37.6\log_{10} (R_{macro})$\\
        \midrule
        Microcell pathloss     &  $140.7+36.7\log_{10} (R_{micro})$\\
        \midrule
        Inter-cell distance     & 0.2 km\\
        \midrule
        User-BS distance         & Uniform Distribution (0, 0.12 km)\\
        \midrule
        $\rho$     & 0.3 \\
        \midrule
        Noise               & -128.1 dBm\\
        \midrule
        Max transmit power\\of macrocell BS           & 12 W\\
        \midrule
        Max transmit power\\of mircocell BS       & 1.2 W\\
        \bottomrule
    \end{tabular}}
\end{table}

According to the data generation process in section~\ref{sect:dnn}, we generate 20,000 training data sets and 2000 testing data sets. To better extract the characteristics of the CNN, we transpose the channel matrix $\mathbb{H}_{3\times 6\times 2}$ to $\mathbb{H}_{6\times 6\times 1}$ as the input of the neural network.
We compare the performance of the CNN-based approach with four other schemes: 1) DNN by using a FC DNN, as specified in Table~\ref{net_dnn}; 2) Benchmark by using the Exhaustive method; 3) RandomPower by randomly generating the power allocation following a uniform distribution; and 4) MaxPower by allocating the maximum transit power of BSs. The latter two schemes serve as heuristic baselines.
\begin{table}[ht]
    \caption{An Overview of Network Configurations and Parameters. \label{net_dnn}}
    \centering
    \footnotesize
    \renewcommand\arraystretch{0}
    \setlength{\tabcolsep}{4.0mm}{
    \begin{tabular}{c c c}
        \toprule
                &DNN  &CNN\\
        \midrule
        \midrule
        Input Layer& $\mathbb{H}_{6\times 6\times 1}$        &$ \mathbb{H}_{6\times 6\times 1}$\\
        \midrule
        Layer1     &Dense 256-ReLU       & Conv2D 6x6x16-ReLU\\
        \midrule
        Layer2     &Dense 256-ReLU       & Conv2D 6x6x16-ReLU\\
        \midrule
        Layer3     &Dense 128-ReLU       & Conv2D 6x6x32-ReLU\\
        \midrule
        Layer4     &Dense 128-ReLU       & Conv2D 6x6x32-ReLU\\
        \midrule
        Layer5     &    -               &Dense 256-ReLU\\
        \midrule
        Layer6     &    -               &Dense 256-ReLU\\
        \midrule
        Layer7     &    -               &Dense 128-ReLU\\
        \midrule
        Output Layer&  8-way softmax    &   8-way softmax \\
            & and 6-way linear &and 6-way linear\\
        \midrule
        Total Parameters & 126478  &401182\\
        \bottomrule
    \end{tabular}}
\end{table}

\subsection{Simulation Results}
\begin{figure}[ht]
    \centering
    \includegraphics[height=5.0cm, width=8.5cm]{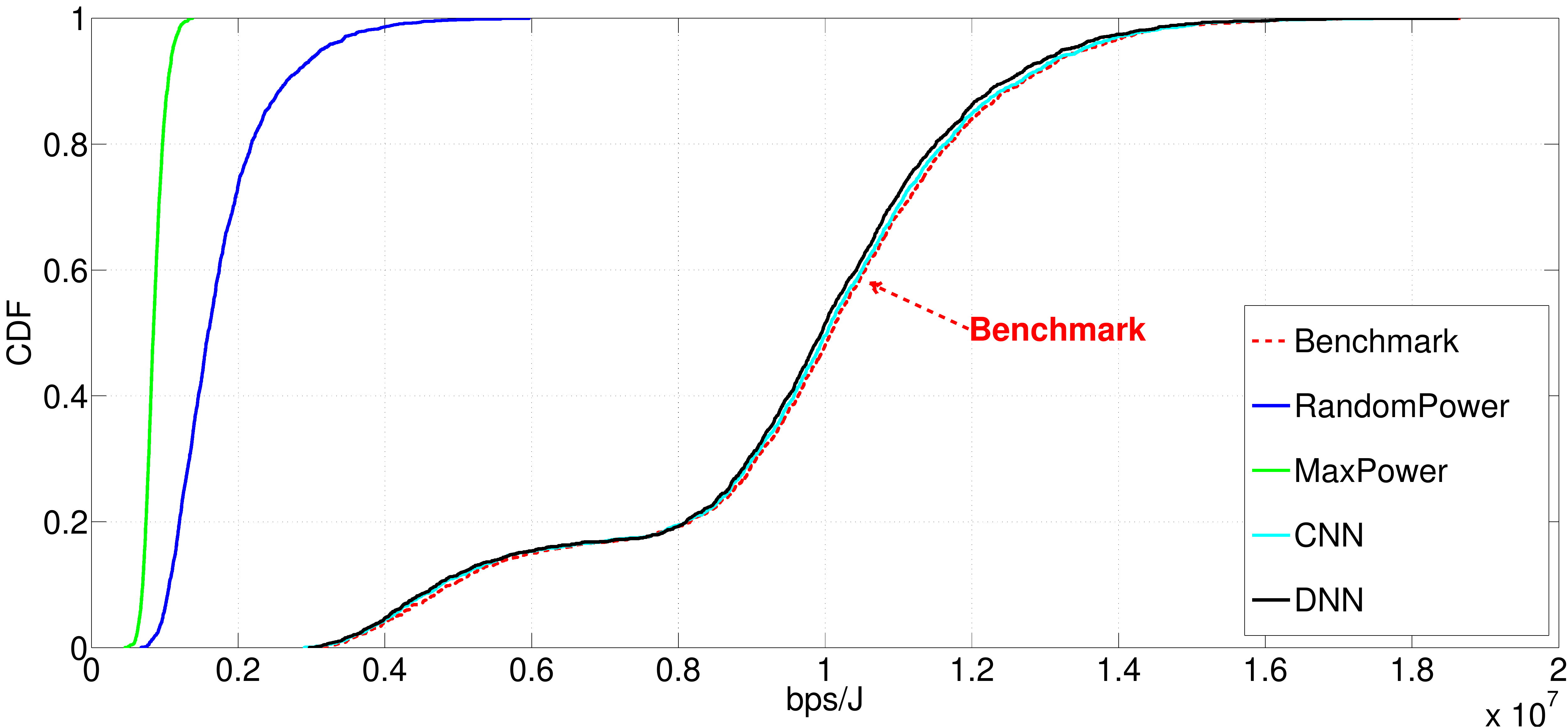}
    \caption{The CDF that describes the EE achieved by different approaches: 1) Benchmark; 2) RandomPower; 3) MaxPower; 4) DNN with 20k training data; 5) CNN with 20k training data.}
    \label{fig:compareMethod}
\end{figure}
\begin{figure}[ht]
    \centering
    \includegraphics[height=5.0cm, width=8.5cm]{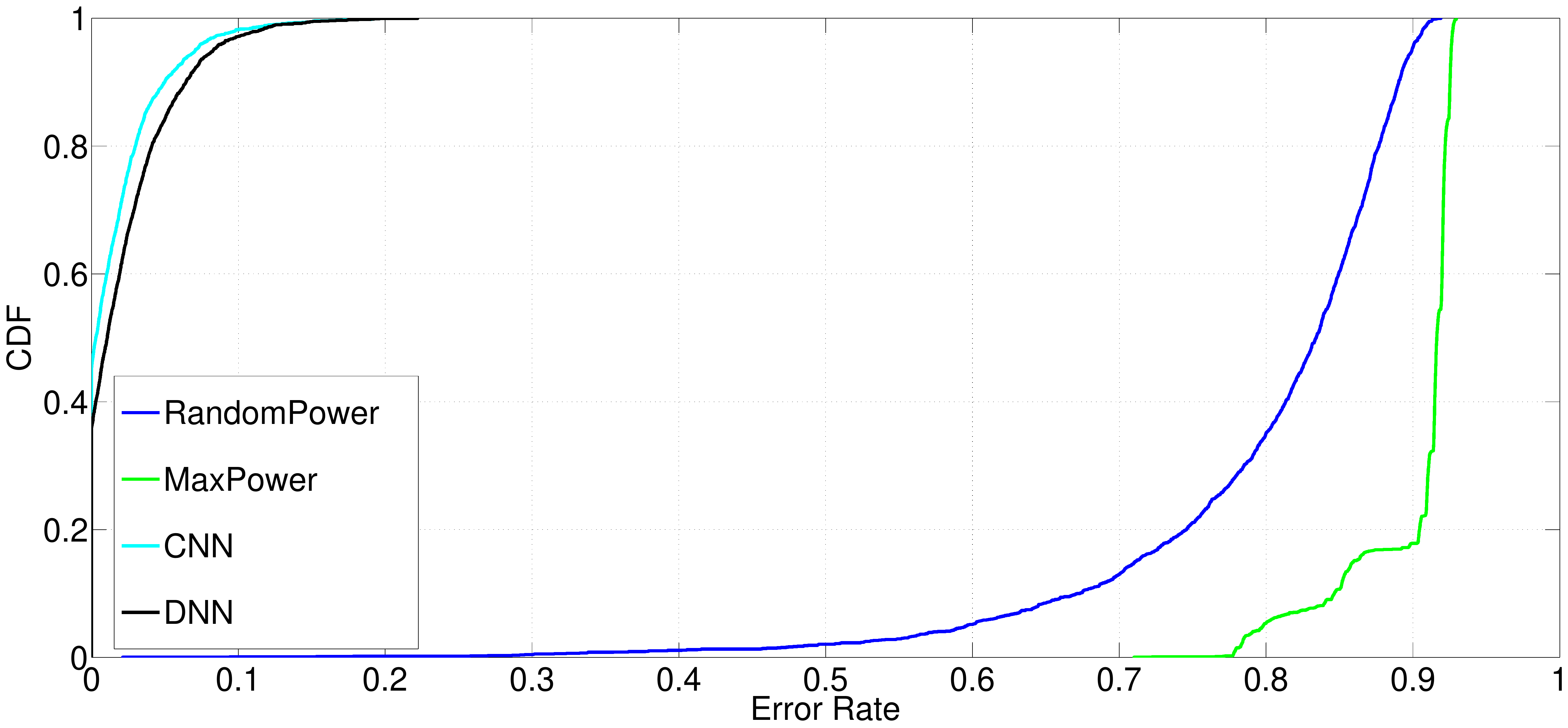}
    \caption{The CDF that describes the error rates of different approaches: 1) RandomPower; 2) MaxPower; 3) DNN with 20k training data; 4)  CNN with 20k training data.}
    \label{fig:CompareError}
\end{figure}

Fig.~\ref{fig:compareMethod} plots the cumulative distribution function (CDF) that describes the EE (in bps/J) achieved by different approaches. As shown in the figure, CNN and DNN can achieve EE very close to Benchmark, while substantially improving the performance of RandomPower and MaxPower.

Fig.~\ref{fig:CompareError} shows the error rates of different approaches compared to Benchmark, defined as
\begin{equation}
       \xi = \frac{|EE_{NN}-EE_{o}|}{EE_{o}}
\end{equation}
where $EE_{NN}$ is the EE of different approaches, and $EE_{o}$ is the EE of Benchmark.
It is observed that CNN incurs the minimum error rate among all approaches, followed by DNN, RandomPower and MaxPower. The error rates  of around 90\% testing data using CNN are lower than 8\% compared to Benchmark.

\begin{figure}[ht]
    \centering
    \includegraphics[height=5.0cm, width=8.5cm]{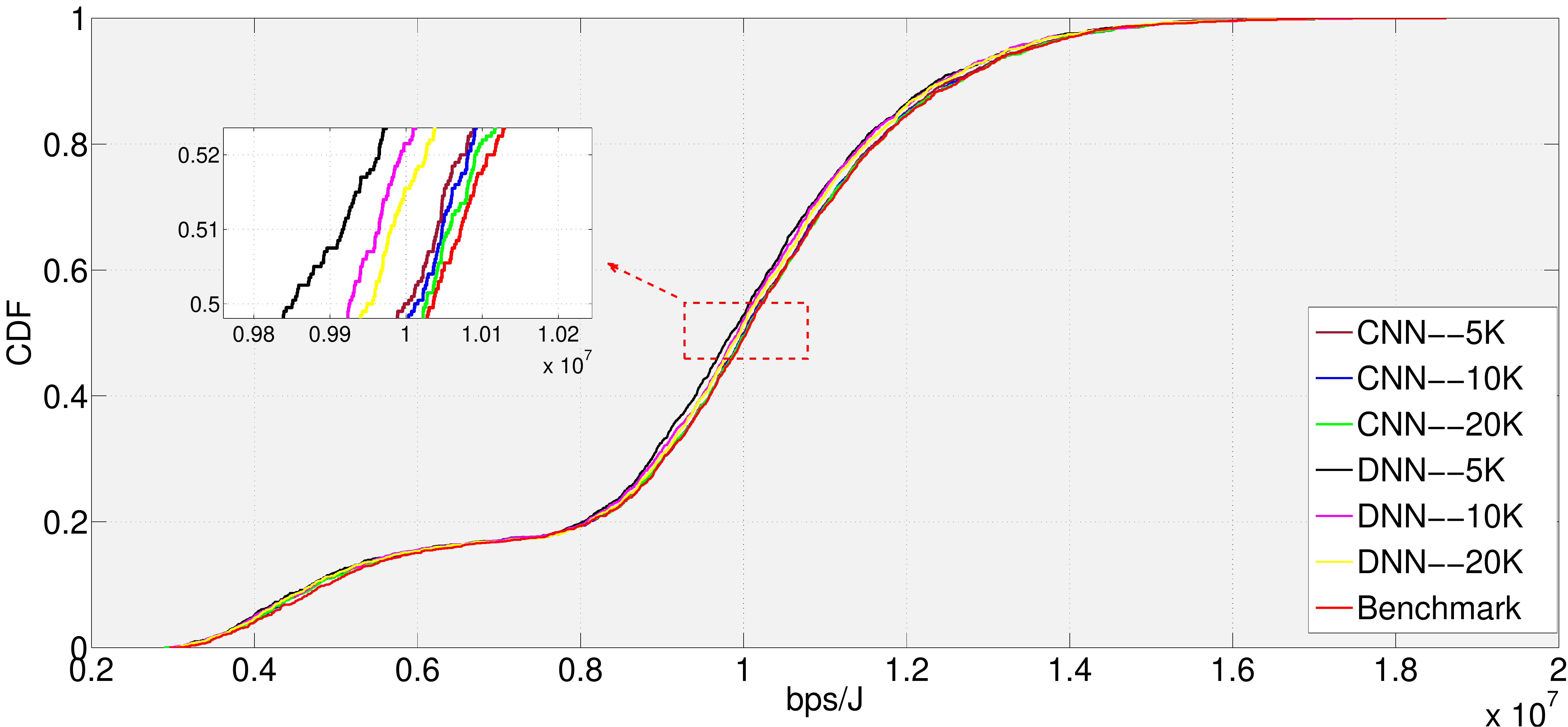}
    \caption{The CDF that describes the EE achieved by the neural networks using different size of training data: 1) CNN-5K; 2) CNN-10K; 3) CNN-20K; 4) DNN-5K; 5) DNN-10K; 6) DNN-20K; and 7) Benchmark.}
    \label{fig:CNN_DNN}
\end{figure}

\begin{figure}[ht]
    \centering
    \includegraphics[height=5.0cm, width=8.5cm]{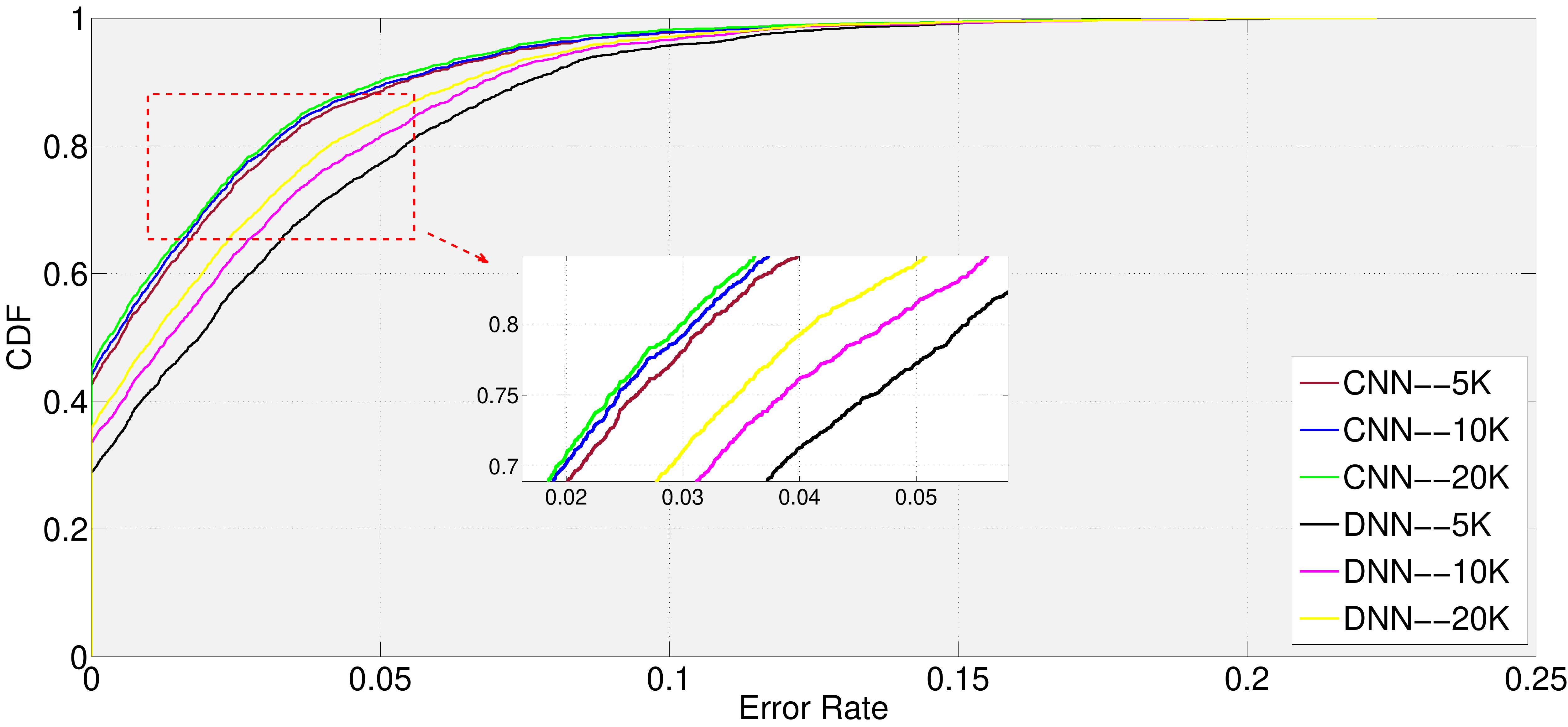}
    \caption{The CDF that describes the error rates of the neural networks using different size of training data: 1) CNN-5K; 2) CNN-10K; 3) CNN-20K; 4) DNN-5K; 5) DNN-10K; and 6) DNN-20K.}
    \label{fig:CNN_DNN_error}
\end{figure}

We also evaluate CNN and DNN with different size of training data. Fig.~\ref{fig:CNN_DNN} and Fig.~\ref{fig:CNN_DNN_error} show the CDFs that describe the EE and error rate of CNN and DNN using different size of training data. We can observe that a network trained with more training data has the performance closer to Benchmark. We can also see that the proposed CNN has better performance than DNN because CNN can extract more detailed data characteristics through sliding windows.


Table~\ref{cputime} lists the CPU runtime of Benchmark, CNN (with 10k or 20k training data), and DNN (with 10k or 20k training data). We can see that the CPU runtime of CNN and DNN with 10k training data is only 6.76\% and 3.94\% of the CPU runtime of Benchmark, respectively. With more parameters in the neural network, the runtime of CNN is slightly bigger than that of DNN. It is also obvious that more training data results in a larger runtime.
\begin{table}[ht]
    \caption{CPU Runtime Comparison \label{cputime}}
    \centering
    \renewcommand{\arraystretch}{2}
    \setlength{\tabcolsep}{1.2mm}{
    \begin{tabular}{|c|c|c|c|c|c|c|}
    \hline
    Method &  Benchmark  &    CNN-20k  & CNN-10k & DNN-20k & DNN-10k \\
    \hline
    Time (s)  &  2.41   & 0.165 & 0.163  & 0.106  & 0.095 \\
    \hline
    $\frac{\text{CNN (DNN)}}{\text{Benchmark}}$ & - & 6.85\% & 6.76\% & 4.4\% & 3.94\% \\
    \hline
    \end{tabular}}
\end{table}

\section{Conclusion} \label{sect:conc}
By introducing deep learning technology to resource allocation problem in wireless communications, we proposed a CNN-based approach to maximize the EE for HetNets. The proposed approach decomposed the original problem into a classification subproblem and a regression subproblem, and output the energy-efficient decisions on joint subchannel and power allocation with a low computational complexity. Extensive numerical experiments demonstrated that the proposed CNN achieved similar performance as the Exhaustive method, while needed only 6.76\% of its CPU runtime.

\section*{Acknowledgment}
This work was supported by the National Natural Science Foundation of China (NSFC) Grants under No. 61701293 and No. 61871262, the National Science and Technology Major Project Grants under No. 2018ZX03001009, the Huawei Innovation Research Program (HIRP), and research funds from Shanghai Institute for Advanced Communication and Data Science (SICS).

\bibliographystyle{IEEEtran}
\bibliography{IEEEfull,ref}

\end{document}